%% file: paper.tex
\documentstyle[11pt,aaspp4,psfig,natbib209]{article}
\input macros
\def\tmodel{{\rm T}_{\rm model}}
\begin{document}
\bibliographystyle{apj}

\title{Detailed NLTE Model Atmospheres for Novae during Outburst:\\
I. New Theoretical Results.}

\author{Peter H. Hauschildt}
\affil{Dept.\ of Physics and Astronomy, 
University of Georgia, Athens, GA 30602-2451\\
Email: {\tt yeti@hal.physast.uga.edu}}
\author{Steven N. Shore}
\affil{Dept.\ of Physics and Astronomy, Indiana University South Bend\\
1700 Mishawaka Ave, South Bend, IN 46634-7111\\
E-Mail: \tt sshore@paladin.iusb.edu}
\author{Greg J. Schwarz}
\affil{Dept.\ of Physics and Astronomy, 
Arizona State University, Tempe, AZ 85287-1504\\
Email: {\tt schwarz@hydro.la.asu.edu}}
\author{E. Baron}
\affil{Dept. of Physics and Astronomy, University of Oklahoma
440 W. Brooks, Rm 131, Norman, OK 73019-0225\\
E-Mail: \tt baron@phyast.nhn.ou.edu}
\author{S. Starrfield}
\affil{Dept.\ of Physics and Astronomy, 
Arizona State University, Tempe, AZ 85287-1504\\
Email: {\tt sumner.starrfield@asu.edu}}
\and
\author{France Allard}
\affil{Dept.\ of Physics, Wichita State University,
Wichita, KS 67260-0032\\
E-Mail: \tt allard@eureka.physics.twsu.edu}

\begin{abstract}

We present new detailed NLTE calculations for model atmospheres of
novae during outburst.  This fully self-consistent NLTE treatment
for a number of model atoms includes 3922 NLTE levels and 47061 NLTE
primary transitions.  We discuss the implication of departures from LTE
for the strengths of the lines in nova spectra. The new results show
that our large set of NLTE lines constitute the majority of the total
line blanketing opacity in nova atmospheres. Although we include LTE
background lines, their effect are small on the model structures and on
the synthetic spectra.  We demonstrate that the assumption of LTE leads to
incorrect synthetic spectra and that NLTE calculations are required for
reliably modeling nova spectra. In addition, we show that detailed NLTE
treatment for a number of ionization stages of iron changes the results
of previous calculations and improve the fit to observed nova spectra.

\end{abstract}

\section{Introduction}

In a series of papers
\cite[]{novapap,cygpap,cas93pap,novaphys,novsim,fe2nova,iau152,osand}
we have developed detailed spherical, expanding NLTE model atmospheres
to treat the optically thick, early stages of nova outbursts and
have also analyzed a number of observed nova spectra. The models were
computed self-consistently, employing the equation of transfer in a
special relativistic framework \cite[]{s3pap}, energy conservation in
the co-moving frame calculated with a modified Uns\"old-Lucy method
\cite[]{MDpap}, NLTE effects for a large number of atomic and molecular
species using numerical methods that we have developed to treat very
large and detailed model atoms \cite[]{casspap,fe2pap,parapap}, and
extensive NLTE and LTE line blanketing by several million lines that
were dynamically selected from the Kurucz lists \cite[]{cdrom1} and
other sources (for molecular lines).  These models have been extremely
successful in fitting observed nova spectra \cite[]{cas93pap,osand},
and, in particular, have been used to identify the UV signature of the
``fireball'' phase of Nova Cygni 1992 \cite[]{cygpap} and LMC~1991
(Schwarz \etal, in preparation). 

Although nova atmospheres can be treated as stationary expanding shells
\cite[cf.][]{bath76}, the time development of the nova spectra can be
simulated by a series of nova atmospheres at constant luminosity and
increasing model temperatures during the phase of constant bolometric
luminosity of the nova. Therefore, the analysis of a time sequence of
observations can provide insight, for instance, into the development of
the velocity field of the shell.  Such sequences can also be used to check
the results of the analysis for consistency and to obtain an estimate
on the statistical errors inherent in the analysis process. This means
that grids of nova atmosphere models and synthetic spectra are important
to investigate the time evolution of a nova outburst.

We present here a new generation of nova atmosphere models and synthetic
spectra with a much larger set of NLTE species than any previous
calculation.  The majority of the line blanketing opacity in these
new models is provided by NLTE lines, which constitutes a significant
advance for nova model atmospheres. In this first paper, we discuss
some new theoretical results and their implications.  In a companion
paper, we show fits to specific observed novae using these models.
Our outline is as follows.  In section 2, we briefly discuss the model
assumptions and the numerical methods that we use and how we derive our
model atoms. The results of the calculations are presented in section 3,
and the structure of the atmosphere and the spectra are discussed in
section 4.  We close with a summary and conclusions.

\section{Methods and Models}

The basic assumptions of our nova models are the same as used in
\cite{novaphys}.  The expanding nova shell is assumed to have a power
law density of the form $\rho\propto r^{-N}$ with $N=3$ for the models
presented in this paper. The velocity law is derived from the
condition of constant mass-loss rate (in radius) with a prescribed maximum
velocity $\vmax$ (we use here $\vmax=2000\kms$), consistent with typical
values observed in classical novae. We further parameterize the models
with the ``model temperature'' $\tmodel$ through the relation $L=4\pi
R^2 \sigma \tmodel^4$, where $L$ is the luminosity of the model (here set
to $L=50,000\Lsun$ for all models, the absolute value of the luminosity
does {\em not} affect the spectra, see \cite{novsim,novaphys}).  $R$
is the radius of the shell at $\tau=1$ in the bound-free (hereafter,
b-f) continuum at $5000\ang$. The model temperature is comparable to
the effective temperature $\teff$, which is well defined only for plane
parallel atmospheres, in that it parameterizes the bolometric flux of
the model at any time. However, it should not be confused with 
the effective temperature,which can only be defined for plane-parallel
configurations \cite[]{novaphys}.

We solve the radiative transfer equation consistently for lines and
continua (allowinf for arbitrary overlaps) with the method discussed
in \cite{s3pap} and \cite{novaphys} rather than employing the the
Sobolev approximation.  \cite{novaphys} have shown that this simpler
method {\it cannot} be used in nova atmospheres due to the large number
of overlapping lines as well as the strong coupling between lines
and continua.  Such complications require that the multi-level NLTE
rate equations be solved self-consistently and simultaneously with the
radiative transfer and energy equations, and the equations must include
the effects of both line blanketing and expansion of the nova atmosphere.

For this analysis, we use our multi-purpose stellar atmosphere code
\phoenix.  \phoenix\ \cite[version 8.0,][]{fe2nova,parapap} uses a
special relativistic spherical radiative transfer for nova models and an
equation of state (EOS) which includes more than 300 ions of 39 elements
(with up to 26 ionization stages each).  The temperature correction is
based on e variety of the Uns\"old-Lucy method that has been modified
to include NLTE and scattering.  This algorithm converges very quickly
and is highly stable.

Both the NLTE and LTE background lines (see below) are treated with
a direct opacity sampling method.  We do {\em not} use pre-computed
opacity sampling tables, but instead dynamically select the relevant
LTE background lines from master line lists at the beginning of each
iteration and sum the contribution of every line within a search window
to compute the total line opacity at {\em arbitrary} wavelength points.
The latter feature is crucial in NLTE calculations in which the wavelength
grid is both irregular and variable from iteration to iteration due to
changes in the physical conditions. This approach also allows detailed
and depth dependent line profiles to be used during the iterations. To
make this method computationally efficient, we employ modern numerical
techniques, e.g., vectorized and parallel block algorithms with high
data locality \cite[]{parapap}, and we use high-end workstations or
supercomputers for the model calculations.

In the calculations we present in this paper, we have have included a
constant statistical velocity field, $\xi = 50\kms$, which is treated like
a microturbulence.  The choice of lines of species not explicitly treated
in NLTE (hereafter, LTE background lines) is dictated by whether they are
stronger than a threshold $\Gamma\equiv \chi_l/\kappa_c=10^{-4}$, where
$\chi_l$ is the extinction coefficient of the line at the line center and
$\kappa_c$ is the local b-f absorption coefficient.  This typically leads
to about $2\alog{6}$ LTE background lines.  The profiles of these lines
are assumed to be depth-dependent Gaussians.  We have verified in test
calculations that the details of the LTE background line profiles and the
threshold $\Gamma$ do not have a significant effect on either the model
structure or the synthetic spectra. However, the LTE background lines
are included because their cumulative effect can change the
structure and the synthetic spectra.  In addition, we include about
2000 photo-ionization cross sections for atoms and ions (Verner \&
Yakovlev 1995)\nocite{verner95}.

Hauschildt \& Baron (1995\nocite{fe2pap}) have extended the numerical
method developed by Hauschildt (1993)\nocite{casspap} for NLTE
calculations with an extremely detailed model atom of Fe~II. In the
calculations presented in this paper, we significantly enlarge the set of
NLTE species, namely H, He~I--II, Mg~II, Ca~II, Ne~I, C~I--IV, N~I--VI,
O~I--VI, S~II--III, Si~II--III, and Fe~I--III (for a complete list
of NLTE species available in \phoenix\ 8.0 see Table \ref{nltetab}).
Note that we do {\em not} use the NLTE treatment for Li~I, Na~I,
Co~I---III and Ti~I--III because these species are not very important
in nova atmospheres and particularly treating Co and Ti in NLTE would
considerably increase the CPU time for the model calculations with
little additional improvements.  We thus include a total of  3922 NLTE
levels and 47061 NLTE primary lines in the calculations presented here,
nearly a factor of 5 more levels and lines than in our previous nova
calculations \cite[]{fe2nova}.

\subsection{NLTE Calculational Method}

The large number of transitions that have to be included in realistic
models of nova atmospheres require an efficient method
for the numerical solution of the multi-level NLTE radiative transfer
and model calculation  problem.  Simple approximations, such as the
Sobolev method, are very inaccurate in situations in which lines overlap
strongly and make a significant pseudo-continuum contribution (weak lines), as is the case for nova and supernova atmospheres
\cite[cf.][]{fe2nova,fe2sn}.  Classical techniques, such as the
complete linearization or the Equivalent Two Level Atom method, are
computationally prohibitive. In addition, we are modeling moving
media (e.g., stellar winds, novae and supernovae), so that approaches
such as Anderson's multi-group scheme \cite[]{anderson89} or extensions
of the opacity distribution function method \cite[]{HubLan95} cannot
be applied because of the velocity dependent coupling of different
wavelengths. Methods that are based on partial linearization schemes and
the use of superlevels tend to be numerically less stable, and frequently
encounter convergence problems because of the highly non-linear and
non-local couplings that dominate these atmospheres.

We use the multi-level NLTE operator splitting method described by
Hauschildt (1993)\nocite{casspap}. This method solves the non-grey,
spherically symmetric, special relativistic equation of radiative
transfer in the co-moving (Lagrangian) frame using the operator splitting
method described in Hauschildt\nocite{s3pap} (1992, see also Cannon
[1973]\nocite{cannon}).  This method has been presented in \cite{fe2pap},
so we do not repeat the detailed description here.  For all primary
NLTE lines, the radiative rates and the ``approximate rate operators''
\cite[]{casspap} are computed and included in the iteration process.
Secondary NLTE lines are included for completeness, but do not affect
either the model structure or the synthetic spectra (in fact, the
model atoms have been explicitly constructed so that all important
lines are primary lines). This method is flexible, can be parallelized
\cite[]{parapap} and, most importantly, leads to stable convergence even
for models for which no good starting estimates are available. The latter
property is an advantage if grids of atmospheres have to be calculated.

\subsection{The model atoms}

To construct most of the model atoms we have included all observed
levels that have observed bound-bound (hereafter, b-b) transitions
with $\log{(gf)} > -3.0$ as NLTE levels where $g$ is the statistical
weight of the lower level and $f$ is the oscillator strength
of the transition (we have used Kurucz's data provided on CD-ROMs;
\cite{cdrom18,cdrom20,cdrom21,cdrom22,cdrom23}). That is, we solve the
complete b-f and b-b radiative transfer and rate equations for all levels
including all radiative rates of the primary lines. 
%Grotrian diagrams
%of these model atoms are shown in Fig.~\ref{C-grot}--\ref{Fe-grot}.
In addition, we treat the opacity and emissivity for the remaining
$\approx 2$ million weak secondary b-b transitions in NLTE if one level
of a secondary transition is included in the model. A more complete
description of the numerical method is presented in Hauschildt \& Baron
(1995)\nocite{fe2pap}.

Photo-ionization and collisional rates for most of the model atoms are
not yet available.  Thus, we have taken the results of the Hartree-Slater 
central field calculations of Reilman \& Manson (1979)\nocite{reilman79}
to scale the ground state photo-ionization rate to the species of interest
and have then used a hydrogenic approximation for the energy 
dependence 
of the cross-section. Although this provides only a rough approximation,
the exact values of the b-f cross-sections are not important for the
opacities themselves, which are dominated by known b-b transitions.

A more accurate treatment of these model atoms requires more accurate
collisional and photo-ionization rates, which are not presently available.
We have, therefore, been forced to make several simplifying choices.
While collisional rates are important in hotter stellar atmospheres
with high electron densities, they remain nearly negligible compared
to the radiative rates for the low electron densities found in cooler
nova ejecta.  We have approximated b-f collisional rates using the
semi-empirical formula of Drawin (1961)\nocite{drawin61}. The b-b
collisional rates were approximated by the semi-empirical formula
of Allen (1973)\nocite{allen_aq}, while Van~Regemorter's formula
(1962\nocite{vr62}) was used for permitted transitions.  In the present
calculations we have neglected collisions with particles other than
electrons because the cross-sections are largely unknown.

\section{Results}

We have computed a grid of models with 50 radial points to investigate the
effects of NLTE on the atmospheric structure and the spectra of novae.
All models include the NLTE treatment as discussed above as well as the
standard \phoenix\ equation of state and additional LTE background lines
(about 2 million atomic lines). The NLTE effects are included in both
the temperature iterations (so that the structure of the models includes
NLTE effects) and all radiative transfer calculations.  For primary
NLTE lines we add 3 to 5 wavelength points within their profiles to the
global wavelength grid (the transfer equation, the rate operators and
the approximate rate operators are computed for every wavelength point
that falls into the profile of each of each primary line, resulting in
substantially more wavelength points per line due to the line crowding).
This procedure typically leads to about 120,000--190,000 wavelength points
for the model and the synthetic spectrum calculations. We iterated the
models until the radiative energy conservation is fulfilled better than
1\% for both the co-moving frame flux and its derivative and the NLTE
departure coefficients are converged better than 1\% at the same time.
We have found in test calculations that this is necessary and sufficient
to obtain a converged solution \cite[see also][]{casspap}.  On a single
processor Cray C90 or a single processor HP 9000-C180 workstation, a
typical model calculation with 10 iterations requires about 18 hours CPU
time. On 5 thin-nodes of an IBM SP2 parallel machine using the parallel
version of \phoenix\ \cite[]{parapap}, the same calculations also require
about 18 hours of wall-clock time.

\subsection{NLTE effects on the concentration of atoms and ions}

We plot the relative concentration, $P_i/\pgas$, where $\pgas$ is the
total gas pressure, as a function of the standard optical depth $\tstd$
(which we define as the absorption optical depth in the continuum at
$5000\ang$) for a variety of atoms and ions for both the LTE and the
NLTE cases in Figs.~\ref{H-He-ion}--\ref{Fe-ion}.  The LTE plots were
constructed using the NLTE structure of the model atmosphere but setting
all departure coefficients, $b_i$, to unity.  Figure \ref{H-He-ion}
shows the ionization balance for H and He as well as the concentration
of free electrons in a nova model with $\tmodel=15000\,$K.  In this
model, hydrogen recombines in the optically thinner regions of the
atmosphere. Therefore, the NLTE effects lead to changes in the location
and size of the recombination zone as compared to the LTE models.
In particular, they lead to an earlier recombination of H~II to H~I
at an optical depth of $\tstd\approx 3\alog{-3}$ compared to the LTE
location of the H recombination zone around $\tstd\approx 10^{-5}$. In
the outer atmosphere, however, the NLTE model shows a higher electron
density compared to the LTE structure. Here the NLTE effects cause a
slight over-ionization of hydrogen. To a lesser degree this is also the
case for He~I.  We remind the reader that any one of these models is 
actually a snapshot of the atmosphere {\it in time}, so the 
differences between the LTE and NLTE models will be important for the 
comparison between theory and observations in a sequence of spectra 
for novae in outburst.

In general, the effects of NLTE on the ionization balance
are relatively small for C, N, and O. We demonstrate this in
Figs.~\ref{C-ion}--\ref{O-ion} where we plot the CNO ionization balance
for those models with the largest NLTE ionization changes for these
species.  For carbon and nitrogen, the dominant ionization stage in the
atmosphere is practically unaffected by NLTE over- or under-ionization.
Less important ionization stages are affected, but their concentration
is small compared to the dominant ionization stage. For oxygen, the
situation is slightly different, as O$^{+}$ recombines to O$^{o}$
in the outer atmosphere at low optical depths.  The location of the
recombination point is now changed by the NLTE effects, such that the
NLTE models show a higher ionization in the outer atmosphere.

The ionization changes are more pronounced for sulfur and silicon;
cf.\ Figs.~\ref{S-ion} and \ref{Si-ion}. Both elements have recombination
zones in the optically thin parts of some nova models, and the location
and the extent of these zones are altered by the NLTE effects. For these
elements, the NLTE effects are not large enough to change the degree
of ionization significantly farther away from ionization/recombination
zones in the nova atmosphere.  The concentrations of the non-dominant
ionization stages are influenced by NLTE effects, but the lines of these
species are much weaker because of the relatively small number density
of their parent ions. For the elements discussed so far, NLTE effects
will be mostly present in the line formation rather than in changing
the number density of the ions themselves.

For iron, the NLTE effects on the ionization balance are small but
significant. We show this in Fig.~\ref{Fe-ion} for a model with
$\tmodel=20000\,$K.  NLTE reduces the concentration of Fe$^{+}$ ions
in the Fe~II line forming region ($10^{-4} \le \tstd \le 10^{-2}$),
thus reducing the overall strength of the Fe~II lines compared to the
LTE case. These results are different from those we reported previously
\cite[]{fe2nova} because, in these new models, we also include NLTE
effects for Fe$^{o}$ and Fe$^{+2}$.  In particular, the new treatment
of Fe~III reduces the NLTE effects of the Fe$^{+}$/Fe$^{+2}$ ionization
balance; the full treatment of Fe~III reduces the NLTE over-ionization
for Fe~II compared to our previous calculations. We show this in
Fig.~\ref{FeII-ion}, where we compare the full NLTE model (large symbols)
with a model where Fe~II is the {\em only} iron species treated in NLTE
(but all other NLTE species are included). This shows that nova atmosphere
calculations require a comprehensive NLTE treatment with many ions of
the NLTE species. The new models fit the observed spectra of novae better
than previous model generations (Schwarz \etal., in preparation).

The behavior of the iron ionization balance becomes clearer by looking at
the departure coefficients for Fe~I--III for the $\tmodel=20000\,$K model
plotted in Fig.~\ref{Fe-bi} \cite[see ][ for a more detailed discussion of
the departure coefficients and source functions]{fe2nova}.  The variations
of $b_i$ with optical depth are complicated, since their definition
explicitly includes the concentrations of both electrons and the ground
state of the next higher ionization stage \cite[cf.][p.~219]{mihalas78}.
It can be seen that the $b_i$'s of Fe~II and Fe~III are about the same
order of magnitude in the outer region of the atmosphere. Therefore,
the ionization balance of Fe~II to Fe~III is comparable to the LTE case,
because the ratio of the ground state departure coefficients $b_1$
is roughly unity. Treating Fe~III in LTE is equivalent to setting the
departure coefficients of Fe~III to one. This would change the ratio of
the Fe~II to Fe~III $b_1$'s by a factor of $10^4$, over-emphasizing the
change in the ionization balance in the case of pure Fe~II NLTE and Fe~III
NLTE. Therefore, in the full Fe~I--III NLTE treatment, the Fe~II lines
are stronger than in models that treat Fe~II as the only NLTE iron ion.

\subsection{NLTE effects on the synthetic spectra}

\subsubsection{Overview}

In Fig.~\ref{nlte-spec} we compare the synthetic spectra of 4
representative models for $800\ang \le \lambda \le 2\mu$. The figure
shows the progressive changes in the spectra with increasing temperature
of the atmosphere.  The ultraviolet region is particularly sensitive
to such changes.  In order to understand which lines are responsible
for this, we show in Figure \ref{nlte-nolte-spec} synthetic spectra
obtained by omitting LTE background lines; only NLTE lines provide line
blanketing. The spectra are very similar to the full spectra shown in
Fig.~\ref{nlte-spec}. This indicates that the NLTE lines provide the
bulk of the line blanketing and that the LTE background lines, even
though numerous, constitute a comparatively small addition to the total
line opacity.  The fact that our detailed NLTE treatment allows us to
handle the majority of the line blanketing in full NLTE is an important
feature of the calculations, and the result that they constitute the
majority of the line opacity is a basic result of this paper.  The LTE
background lines are, however, nonetheless still important in localized
spectral regions and must be included for the most accurate model and
synthetic spectrum calculations.

An additional result is that the NLTE effects on the synthetic
spectra are far greater than those of the LTE background lines. In
Fig.~\ref{lte-spec} we show the synthetic spectra obtained by setting all
departure coefficients to unity, i.e., the pure LTE assumption, and in
Fig.~\ref{lte2-spec} we show synthetic spectra obtained by using a single
scattering line albedo of 0.95 for all lines, {\it i.e.}, the NLTE lines
are artificially treated in the same manner as the LTE background lines.
An inspection of the figures shows that these synthetic spectra are very
different from the full NLTE results shown in Fig.~\ref{nlte-spec}. This
is particularly true for the infrared lines, which are much stronger
in the pure-LTE spectra than in the NLTE spectra.  These also change
from P Cygni profiles in the NLTE spectra to pure emission lines in
the pure-LTE spectra.  The scattering-LTE infrared spectra shown in
Fig.~\ref{lte2-spec} retain the P Cygni profile but the lines are now
weaker than for the NLTE case. For the IR lines, the NLTE case thus
falls somewhere in between the pure-LTE and scattering-LTE cases.

The IR lines are mainly isolated, single transitions.  The situation is
{\em very different} in the UV, where the lines overlap strongly due to
the level distributions in the iron peak ions and form a pseudo-continuum.
Individual lines are frequently so blended that is impossible to
assign a single identification to a feature in the synthetic spectrum
\cite[see][for discussion of this effect]{novaphys}. Although one
might think that the strong line overlap would reduce NLTE effects,
the computations show that the LTE and NLTE UV spectra are very
different for all models that we have calculated. This can be seen
by closer inspection of Figs.~\ref{nlte-spec}, \ref{lte-spec}, and
\ref{lte2-spec}. In both the pure-LTE and the scattering-LTE spectra,
the UV flux blocking is more severe than in the NLTE models. For $\tmodel
\le 20000\,$K, the LTE spectra show in some regions (blueward of about
$1500\ang$) more than a factor of 10 less flux than the NLTE spectra.
Whereas for $\tmodel=15000\,$K the pure-LTE and scattering-LTE spectra
are similar to each other, the pure-LTE spectra show enormous emission
lines\footnote{Note that these are real emission features, not gaps in
the pseudo-continuum.} for larger model temperatures ($\approx$ 23,000 to
30,000 K) in the $2000\ang$ to $3000\ang$ wavelength range that are absent
in the NLTE spectra. These features are caused by clusters of iron lines
(mostly Fe~II) and the Mg~II h+k doublet. These lines are much weaker in
the scattering-LTE spectra, and they are nearly completely suppressed
in the NLTE spectrum.  This is extremely important for the evaluation
of the optical taxonomy of nova spectra \cite[see][]{williams91}, since
the ``Fe II'' classification is based on the appearance of the optical
emission features.

\subsubsection{1200--$3200\ang$: The IUE spectral region}

In Fig.~\ref{IUE-spec} we show synthetic spectra for models with
$\tmodel=15000\,$K and $30000\,$K in the spectral range of the
International Ultraviolet Explorer satellite (IUE).  The spectra have
been boxcar-smoothed to mimic the IUE low resolution mode (R = 300).
The LTE spectrum shows much less flux than the NLTE model in the short
wavelength end of the range (see below for the same effect in the far UV)
for the $15000\,$K model, which is typical of the spectral appearance at
optical maximum during outburst.  There are a number of features that
are different between the NLTE and either of the two LTE spectra. This
difference is even more pronounced in the model with $\tmodel=30000\,$K,
which is closer to the initial ``fireball'' spectrum and also to
the appearance of novae near UV maximum during the optical decline.
In particular the pure-LTE spectrum shows very strong true emission
lines that are absent in the NLTE spectrum. Below about $2000\ang$ both
LTE spectra show less flux than the NLTE spectrum.  A large fraction of
the absolute flux difference can be traced to the NLTE treatment of CNO
in the full NLTE models and thus shows that a detailed NLTE treatment
of CNO is important for the short wavelength band.

The results here have a bearing on the interpretation of the energy budget
of the atmosphere during the expansion.  Since novae initially evolve at
constant bolometric luminosity, the changes we note between different
treatments of the radiative transfer affect the interpretation of the
physical conditions in the ejecta from observations in relatively small
wavelength intervals.  It is especially important for the determination
of the mass of the white dwarf that the luminosity be accurately obtained
from models, and abundance determinations from the optically thick stage
of the expansion are also dependent upon the correct assessment of the
model temperature and other physical conditions within the atmosphere.

The shape of the pseudo-continuum is comparable between the different
treatments, as might be expected with the enormous number of overlapping
lines (which are essentially the same lines regardless if LTE or NLTE
models are used) in a moving medium. The latter effect also points
out an essential difference between static and moving atmospheres. The
line overlap in moving media is enhanced, and depends on the absolute
velocities as well as on the form of the velocity profile, so that NLTE
versus LTE differences for relatively weak lines are ``washed out''
in moving media, while these differences might very well detectable in
static atmospheres. The pseudo-continuum thus gives information about
the form of the velocity profile if analyzed in detail.

\subsubsection{The region around the Mg~II 2800\AA\ lines}

In Fig.~\ref{Mg-h+k-spec} we show the region around the Mg~II h+k lines
for a nova model with $\tmodel=25000\,$K.  Since this region is well
observed at relatively high resolution (R = 10000 with IUE) during the
early stages of many nova outbursts, it is useful to calculate models
that can be directly compared with the available data.  The upper
panel compares the NLTE spectrum with the pure-LTE and scattering-LTE
spectra. Clearly, the pure LTE assumption is very poor. The scattering-LTE
and NLTE spectra are similar in small wavelength ranges (i.e, between
2430 and $2480\ang$), but there are some significant departures.
The emission component of the Mg~II h+k doublet is similar, but the
absorption trough of the Mg~II lines is too redshifted compared to the
NLTE spectrum.  This shows that, in NLTE, the Mg~II line forming region
has shifted to the outer regions of the nova atmosphere.  The pure-LTE
model shows about the same wavelength shift as the scattering-LTE model
but at a much higher flux level.

The bottom panel of Fig.~\ref{Mg-h+k-spec} compares the full NLTE
spectrum with an NLTE spectrum with {\em no} LTE background lines (dotted
curve). They are very similar, again illustrating that our set of NLTE
species and lines is comprehensive enough to nearly completely describe
the total opacity due to line blanketing. This panel also displays the
result for a model where Fe~I and III were kept in LTE (their lines being
handled like LTE background lines). This is the spectrum for the model
shown in Fig.~\ref{FeII-ion}.  Clearly, the LTE assumption for Fe~I and,
more important for this model, Fe~III produces a significantly different
spectrum.  The total line blanketing due to Fe~II lines is reduced due
to the stronger ionization of Fe~II to Fe~III (see Fig.~\ref{FeII-ion}),
which results in a slightly higher flux level. The Mg~II h+k absorption
trough is stronger and more blue-shifted than in the full NLTE spectrum.
This indicates a significant dependence of the Mg~II line formation on
Fe NLTE caused by line and continua overlaps and structure changes of
the atmosphere in the line forming regions for both Mg~II h+k and the
UV Fe~II lines.

\subsubsection{Near infrared spectra}

In Fig.~\ref{IR-spec} we show the near IR spectra (between $0.83\,\mu$
and $1.1\,\mu$) for two model temperatures ($15000\,$K in the top
panel, $30000\,$K in the bottom panel).  Although this part of the
spectrum has not been extensively studied in most historical outbursts,
it is becoming more accessible at high resolution \cite[see, for
instance][]{1996ApJ...469..854H}.  The infrared becomes less opaque
far earlier in the outburst than any other wavelength region, and is
therefore sensitive to the most rapidly moving parts of the ejecta,
it is important for studying the initial stages of the expansion
and such features as the homogeneity of the ejecta.  In addition,
since it is sensitive to the outer parts of the atmosphere, it may be
less affected by any subsequent wind formation during the outburst.
The scattering-LTE spectrum was scaled so that its continuum coincides
with the NLTE spectrum\footnote{This was necessary because the NLTE
structure was used to generate the LTE spectra.}. The plots show that
the pure-LTE spectra possess much stronger emission lines than the NLTE
models.  This is especially true near $\lambda\approx 8500\ang$, around
the Ca II triplet, for the $\tmodel=15000\,$K model.  The pure-LTE lines
are actually about a factor of 4 {\it stronger} than the NLTE lines in
this region and are off the scale.  In contrast, the scattering-LTE lines
are generally weaker than the NLTE lines in the $\tmodel=15000\,$K model
(the exceptions are two lines at about $8500-8600\ang$), but are stronger
than the NLTE lines in the $30000\,$K model.  The $\tmodel=15000\,$K NLTE
spectrum shows some lines around $9100\ang$ that are in neither of the
LTE spectra.  In the same region, however, the scattering-LTE model shows
lines in the $\tmodel=30000\,$K model. There is no easily identifiable
correspondence between the two LTE spectra and the NLTE spectrum, but the
pure-LTE and scattering-LTE cases seems to bracket the NLTE spectrum for
most lines \cite[see also][for a similar result in SN atmospheres]{fe2sn}.

\subsubsection{Far UV spectra}

The situation in the far UV (FUV) spectral range is very different
from the near IR, cf.\ Fig.~\ref{FUV-spec}.  This region is important
for the energetics of the ``fireball'' and also for the stages just at
and following UV maximum following optical peak.  Although there are
currently no spectra available in this region, with the exception of some
low resolution Voyager 2 data for V1974 Cyg \cite[]{shorecyg94}, this
wavelength region will be accessible once FUSE is launched.  We show the
same models as in Fig.~\ref{IR-spec} and add the NLTE spectra where all
LTE background lines have been omitted (these spectra would have been
identical to the full NLTE spectra shown in Fig.~\ref{IR-spec}). Now
both LTE spectra show less flux than the NLTE spectra. The line overlap
in this spectral region is so severe that all of the features in the
$\tmodel=15000\,$K model, and most of the features in the $30000\,$K
model, are blends.  The LTE background lines have only small effects on
the NLTE spectra, although they produce some absorption features just
above the Lyman edge in the $30000\,$K model.  Again, practically all
of the line blanketing is provided by lines that are treated in NLTE.

\subsubsection{Optical spectra}

We show optical spectra in Figs.~\ref{OPT1-spec} and \ref{OPT2-spec}.
Note that in these plots we plot $F_\nu$ instead of $F_\lambda$ in
order to ``flatten'' the slope of the continuum. Figure \ref{OPT1-spec}
demonstrates that the formation of the Balmer and optical Fe~II lines
{\it cannot} be properly calculated by assuming LTE.  The Fe~II lines
are sensitive to NLTE effects and the pure-LTE and scattering-LTE
approximations bracket the true NLTE shape of the emission parts of
the lines. In addition, their P Cygni profiles cannot be correctly
reproduced by LTE models.  In the $\tmodel=30000\,$K model, the optical
Fe~II lines are far too strong in the LTE spectra compared to the NLTE
spectra. Figure~\ref{OPT2-spec} shows the region around H$\alpha$. The
H$\alpha$ profile is very different between the LTE and NLTE spectra.
While the NLTE emission strength is intermediate between the two LTE
cases, the absorption part of the P Cygni profile is much stronger in
the NLTE spectra than in either of the LTE spectra.  The LTE background
lines have no effect on the optical spectra --- all the detectable lines
in the NLTE spectrum are NLTE lines. This is similar to the case of the
near-IR spectrum.

Thus, pure-LTE is an extremely poor assumption in nova atmospheres,
although scattering-LTE is a somewhat better description. However, {\em
neither approach can replace a detailed and complete NLTE calculation}
--- the complexity of the line formation in a nova atmosphere can only be
adequately described by these NLTE model calculations.

\section{Summary and Conclusions}

In this paper, we have presented a new set of model atmospheres for novae
during the early phase of the outburst, with significantly improved input
physics. Modern numerical methods that allow the detailed treatment of
thousands of NLTE levels make it now possible to include the majority
of the line blanketing opacity in detailed NLTE.  LTE background lines
from species that we do not treat explicitly in NLTE are not important for
the structure of the atmosphere.  They do, however, contribute some of 
the opacity and are necessary in order to reproduce
some of the details in the synthetic spectra, particularly in the UV.

We have discussed the synthetic spectra for a number of models in
detail. The comparison with LTE and simplified NLTE models for a number of
specific wavelength ranges shows that (unfortunately) the detailed models
cannot easily be approximated by simplified approaches.  
The addition of a detailed NLTE treatment for Fe~I and (more importantly)
Fe~III has changed the synthetic spectra compared to our previous results
\cite[]{fe2nova}, which in turn improves the fits to observed nova spectra
(Schwarz \etal, in preparation). Therefore, it is important to include
a number of ionization stages in detailed NLTE to correctly model the
line strengths. 

Our models span a wide range of (model) temperatures. These correspond to
different times of the evolution of novae: In the early wind phase we have
$\tmodel\approx 10000$--$15000\,$K, while in the later pre-nebular stage
$\tmodel\approx 30000\,$K. Thus the analysis of a time sequence of nova
spectra can be used to reduce the error bars of, e.g., abundances and to
check for the internal consistency of the solution. In the second paper
of this series (Schwarz \etal, in preparation), we will present fits to
observed nova spectra obtained with the models presented in this paper.

\smallskip
\begin{small}
\noindent{\em Acknowledgments:}
We thank S. Pistinner for helpful discussions and the referee for very
helpful comments.  This work was supported in part by NASA ATP grant NAG
5-3018 and LTSA grant NAG 5-3619 to the University of Georgia, by NASA
LTSA grants NAGW 4510 and NAGW 2628 and NASA ATP grant NAG 5-3067 to
Arizona State University, STScI grant GO 6082 to IUSB, and by NSF grant
AST-9417242, NASA grant NAG5-3505 and an IBM SUR grant to the University
of Oklahoma.  Some of the calculations presented in this paper were
performed on the IBM SP2 of the UGA UCNS, at the San Diego Supercomputer
Center (SDSC), the Cornell Theory Center (CTC), and at the National Center
for Supercomputing Applications (NCSA), with support from the National
Science Foundation, and at the NERSC with support from the DoE. We thank
all these institutions for a generous allocation of computer time.
\end{small}

\clearpage
\bibliography{yeti,opacity,novae,ltsa,mdwarf,radtran,opacity-fa}

\clearpage
\section{Tables}

\begin{table}
\caption[]{\label{nltetab}Complete listing of {\tt PHOENIX} (version 8)
NLTE species.  The tables entries are of the form N/L, where N is the
number of NLTE levels and L the number of primary NLTE lines for each
model atom. The calculations presented in this paper include all species 
listed in the table with the exceptions of Li, Ti, and Co.}
\smallskip
\begin{tabular}{*{7}{l}}
\hline
\hline
        &   I  & II & III & IV & V & VI \\
\hline
H       &  30/435\\
He      &  11/14 & 15/105\\
Li      &  57/333 \\
C       &  228/1387 & 85/336 & 79/365 & 35/171 \\
N       &  252/2313 & 152/1110 & 87/266 & 80/388 & 39/206 &15/23 \\
O       &  36/66 & 171/1304 & 137/766 & 134/415 & 97/452 & 39/196 \\
Ne      &  26/37 \\
Na      &  3/2 \\
Mg      &  &  18/37 \\
Si      &  & 93/436 & 155/1027 \\
S       &  & 84/444 & 41/170 \\
Ca      &  & 87/455 \\
Ti      & 395/5279 & 204/2399 \\
Fe      & 494/6903 & 617/13675 & 566/9721\\
Co      & 316/4428 & 255/2725 & 213/2248\\
\hline
total:  & & 5346/60637 \\
\hline
\hline
\end{tabular}
\end{table}

\clearpage
\section{Figures}

\begin{figure}[t]
\caption[]{\label{H-He-ion} Ionization balance for hydrogen and helium for a nova
model with $\tmodel=15000\,$K. The plot shows the relative concentration
$P_i/\pgas$. The large symbols show the NLTE model,
the small symbols show the LTE results.}
\end{figure}

\begin{figure}[t]
\caption[]{\label{C-ion} Ionization balance for carbon for a nova
model with $\tmodel=25000\,$K. The plot shows the relative concentration
$P_i/\pgas$ for some carbon ions. The large symbols show the NLTE model,
the small symbols show the LTE results.}
\end{figure}

\begin{figure}[t]
\caption[]{\label{N-ion} Ionization balance for nitrogen for a nova
model with $\tmodel=25000\,$K. The plot shows the relative concentration
$P_i/\pgas$ for some nitrogen ions. The large symbols show the NLTE model,
the small symbols show the LTE results.}
\end{figure}

\begin{figure}[t]
\caption[]{\label{O-ion} Ionization balance for oxygen for a nova
model with $\tmodel=15000\,$K. The plot shows the relative concentration
$P_i/\pgas$ for some oxygen ions. The large symbols show the NLTE model,
the small symbols show the LTE results.}
\end{figure}

\begin{figure}[t]
\caption[]{\label{S-ion} Ionization balance for sulfur for a nova
model with $\tmodel=30000\,$K. The plot shows the relative concentration
$P_i/\pgas$ for some sulfur ions. The large symbols show the NLTE model,
the small symbols show the LTE results.}
\end{figure}

\begin{figure}[t]
\caption[]{\label{Si-ion} Ionization balance for silicon for a nova
model with $\tmodel=25000\,$K. The plot shows the relative concentration
$P_i/\pgas$ for some silicon ions. The large symbols show the NLTE model,
the small symbols show the LTE results.}
\end{figure}

\begin{figure}[t]
\caption[]{\label{Fe-ion} Ionization balance for iron for a nova
model with $\tmodel=20000\,$K. The plot shows the relative concentration
$P_i/\pgas$ for some iron ions. The large symbols show the NLTE model,
the small symbols show the LTE results.}
\end{figure}

\begin{figure}[t]
\caption[]{\label{FeII-ion} Ionization balance for iron for a nova
model with $\tmodel=20000\,$K. The plot shows the relative concentration
$P_i/\pgas$ for some iron ions. The large symbols show a NLTE model where
Fe~I--III are treated in NLTE, the small symbols show the results for
a model where Fe~II is the only Fe ion treated in NLTE. All other NLTE
species are identical for both models.} 
\end{figure}

\begin{figure}[t]
\caption[]{\label{Fe-bi} Departure coefficients for Fe~I--III
for a nova model with $\tmodel=20000\,$K. The $+$ signs indicates the
ground state.}
\end{figure}

\begin{figure}[t]
\caption[]{\label{nlte-spec} Comparison of NLTE synthetic spectra
for the nova models. The dotted line gives the Planck-function corresponding
to the models model temperature.}
\end{figure}

\begin{figure}[t]
\caption[]{\label{nlte-nolte-spec} Comparison of NLTE synthetic spectra
where the all LTE background lines have been neglected.  The dotted
line gives the Planck-function corresponding to the models model
temperature.}
\end{figure}

\begin{figure}[t]
\caption[]{\label{lte-spec} Comparison of synthetic spectra where the all
NLTE departure coefficients have been set to unity. This corresponds to
the pure LTE assumptions for the lines that are normally treated in NLTE.
The dotted line gives the Planck-function corresponding to the models
model temperature.}
\end{figure}

\begin{figure}[t]
\caption[]{\label{lte2-spec} Comparison of synthetic spectra where the
all NLTE departure coefficients have been set to unity and the lines
are assumed to have an albedo for single scattering of $0.95$. This
corresponds to the assumptions for the LTE background lines.  The dotted
line gives the Planck-function corresponding to the models model
temperature.}
\end{figure}

\begin{figure}[t]
\caption[]{\label{IUE-spec} Comparison of synthetic spectra for
nova model atmospheres with $\tmodel=15000\,$K and $\tmodel=30000\,$K
in the IUE wavelength range.  The spectra have been boxcar smoothed
to the low resolution of the IUE satellite ($\simeq 6\ang$).  In the
``pure-LTE'' spectrum all NLTE departure coefficients have been set to
unity whereas in the ``scattering-LTE'' spectrum the lines are assumed
to have, in addition, an albedo for single scattering of $0.95$. The
latter corresponds to the assumptions for the LTE background lines.
All spectra use the structure of the full NLTE model.}
\end{figure}

\begin{figure}[t]
\caption[]{\label{Mg-h+k-spec} Comparison of synthetic spectra for nova
model atmospheres with $\tmodel=25000\,$K. In the ``pure-LTE'' spectrum
all NLTE departure coefficients have been set to unity whereas in the
``scattering-LTE'' spectrum the lines are assumed to have, in addition,
an albedo for single scattering of $0.95$. The latter corresponds to the
assumptions for the LTE background lines.  In the ``NLTE, no LTE lines''
spectrum, all LTE background lines have been neglected. The ``NLTE, LTE
for Fe~I and III'' spectrum uses LTE for Fe~I and III (with an albedo
for single scattering of $0.95$) but includes all other NLTE species, in
particular Fe~II). This model has been fully iterated with its set of NLTE
species.  All other spectra use the structure of the full NLTE model.}
\end{figure}

\begin{figure}[t]
\caption[]{\label{IR-spec} Comparison of synthetic spectra for nova
model atmospheres with $\tmodel=15000\,$K and $\tmodel=30000\,$K in the near
infrared. In the ``pure-LTE'' spectrum all NLTE departure coefficients
have been set to unity whereas in the ``scattering-LTE'' spectrum the
lines are assumed to have, in addition, an albedo for single scattering of
$0.95$. The latter corresponds to the assumptions for the LTE background
lines. The scattering-LTE spectra have been scaled so that their continua
coincide with the NLTE continua. All spectra use the structure of the
full NLTE model.}
\end{figure}

\begin{figure}[t]
\caption[]{\label{FUV-spec} Comparison of synthetic spectra for nova
model atmospheres with $\tmodel=15000\,$K and $\tmodel=30000\,$K in the 
far Ultraviolet (FUV).  
In the ``pure-LTE'' spectrum all NLTE departure coefficients
have been set to unity whereas in the ``scattering-LTE'' spectrum the
lines are assumed to have, in addition, an albedo for single scattering of
$0.95$. The latter corresponds to the assumptions for the LTE background
lines. The spectrum labeled ``NLTE, no LTE lines'' is identical to the NLTE
spectrum but omits LTE background lines. All spectra use the structure of the
full NLTE model.}
\end{figure}

\begin{figure}[t]
\caption[]{\label{OPT1-spec} Comparison of synthetic spectra for nova
model atmospheres with $\tmodel=15000\,$K and $\tmodel=30000\,$K in the 
optical. In the ``pure-LTE'' spectrum all NLTE departure coefficients
have been set to unity whereas in the ``scattering-LTE'' spectrum the
lines are assumed to have, in addition, an albedo for single scattering of
$0.95$. The latter corresponds to the assumptions for the LTE background
lines. The spectrum labeled ``NLTE, no LTE lines'' is identical to the NLTE
spectrum but omits LTE background lines. All spectra use the structure of the
full NLTE model.}
\end{figure}
\clearpage
\begin{figure}[t]
\caption[]{\label{OPT2-spec} Comparison of synthetic spectra for nova
model atmospheres with $\tmodel=15000\,$K and $\tmodel=30000\,$K around
H$\alpha$. In the ``pure-LTE'' spectrum all NLTE departure coefficients
have been set to unity whereas in the ``scattering-LTE'' spectrum the
lines are assumed to have, in addition, an albedo for single scattering of
$0.95$. The latter corresponds to the assumptions for the LTE background
lines. The spectrum labeled ``NLTE, no LTE lines'' is identical to the NLTE
spectrum but omits LTE background lines. All spectra use the structure of the
full NLTE model.}
\end{figure}

\end{document}

%% file: macros.tex
\def\la{\mathrel{\hbox{\rlap{\hbox{\lower4pt\hbox{$\sim$}}}\hbox{$<$}}}}
\def\ga{\mathrel{\hbox{\rlap{\hbox{\lower4pt\hbox{$\sim$}}}\hbox{$>$}}}}

\newcommand{\be}{\begin{eqnarray}}
\newcommand{\ee}{\end{eqnarray}}

\newcommand{\msol}{\ifmmode{{\rm M}_\odot}\else{M$_\odot$}\fi}
\newcommand{\foe}{\ifmmode{10^{51}}\else{$10^{51}$}\fi}

\newcommand{\xni}{\ifmmode{{\rm X}_{\rm Ni}}\else{X$_{\rm Ni}$}\fi}
\def\ang{\hbox{\AA}}
\def\Teff{\ifmmode{T_{\rm eff}}\else{\hbox{$T_{\rm eff}$} }\fi}
\def\Rzero{\ifmmode{R_0}\else{\hbox{$R_0$} }\fi}

\def\SP2{{\tt IBM SP2}}

\def\PC2{{\tt PC$^2$}}

\def\inu{\ifmmode{I_{\nu}}\else{\hbox{$I_{\nu}$} }\fi}
\def\snu{\ifmmode{S_{\nu}}\else{\hbox{$S_{\nu}$} }\fi}
\def\jnu{\ifmmode{J_{\nu}}\else{\hbox{$J_{\nu}$} }\fi}

\def\etal{et al.}
\def\fep{\ifmmode{{\rm Fe II}}\else\hbox{Fe~II }\fi}

  at 12.0 true pt %scaled \magstep1

\def\etal{{et al}}

\def\phoenix{{\tt PHOENIX}}

\def\etal{{et al}}

\def\phoenix{{\tt PHOENIX}}

\def\b{\beta}

\def\rout{\ifmmode{r_{\rm out}}\else\hbox{$r_{\rm out}$}\fi}
\def\tmax{\ifmmode{\tau_{\rm max}}\else\hbox{$\tau_{\rm max}$}\fi}
\def\tstd{\ifmmode{\tau_{\rm std}}\else\hbox{$\tau_{\rm std}$}\fi}
\def\vmax{\ifmmode{v_{\rm max}}\else\hbox{$v_{\rm max}$}\fi}
\def\muE{\ifmmode{\mu_{\rm E}}\else\hbox{$\mu_{\rm E}$}\fi} 
\def\pE{\ifmmode{p_{\rm E}}\else\hbox{$p_{\rm E}$}\fi} 
\def\bmax{\ifmmode{\b_{\rm max}}\else\hbox{$\b_{\rm max}$}\fi}
\def\kms{\hbox{$\,$km$\,$s$^{-1}$}}

\def\ang{\hbox{\AA}}

\def\Lsun{\hbox{$\,$L$_\odot$} }
\def\Teff{\hbox{$\,T_{\rm eff}$} }

\def\alog#1{\times 10^{#1}}

\def\rout{\hbox{$r_{\rm out}$} }
\def\pgas{\hbox{$P_{\rm gas}$} }
\def\chistd{\ifmmode{\chi_{\rm std}}\else\hbox{$\chi_{\rm std}$}\fi}

\def\msol{$M_\odot$}
\def\foe{10^{51}}

\def\xni{{\rm X}_{\rm Ni}}

\def\teff{{\rm T}_{\rm eff}}

\def\lstar{\ifmmode{\Lambda^*}\else\hbox{$\Lambda^*$}\fi} 
\def\Rop{\ifmmode{[R_{ij}]}\else\hbox{$[R_{ij}]$}\fi}

\def\Rji{\ifmmode{[R_{ji}]}\else\hbox{$[R_{ji}]$}\fi}
\def\Rstar{\ifmmode{[R_{ij}^*]}\else\hbox{$[R_{ij}^*]$}\fi}

\def\Rjistar{\ifmmode{[R_{ji}^*]}\else\hbox{$[R_{ji}^*]$}\fi}
\def\DRji{\ifmmode{[\Delta R_{ji}]}\else\hbox{$[\Delta R_{ji}]$}\fi}
\def\DRij{\ifmmode{[\Delta R_{ij}]}\else\hbox{$[\Delta R_{ij}]$}\fi}

\def\ns{\ifmmode{N_{\rm s}}          % Anzahl der tau-punkte
        \else\hbox{$N_{\rm s}$}\fi}

\def\mat#1{{\bf #1}}     % Macro fr Matrizen
\def\vek#1{{#1}}         % Macro fr Vektoren

\newcount\eqcount
\def
  \nummer{
    \global\advance\eqcount by 1
    (\the\eqcount)
  }

\def
  \numadv{
    \global\advance\eqcount by 1
  }

\def
   \numout#1{
     (\the\eqcount #1)
  }

\def\ivek#1#2{\ifmmode{\vek{I}^{#1}_{#2}}
        \else\hbox{$\vek{I}^{#1}_{#2}$}\fi}

\def\tmat#1#2{\ifmmode{\mat{t}^{#1}_{#2}}
        \else\hbox{$\mat{t}^{#1}_{#2}$}\fi}
\def\rmat#1#2{\ifmmode{\mat{r}^{#1}_{#2}}
        \else\hbox{$\mat{r}^{#1}_{#2}$}\fi}
\def\bvek#1#2{\ifmmode{\beta^{#1}_{#2}}
        \else\hbox{$\beta^{#1}_{#2}$}\fi}

\def\lp{\ifmmode{\lambda^+_\tau}           % lambda +
        \else\hbox{$\lambda^+_\tau$}\fi}
\def\lm{\ifmmode\lambda^-_\tau             % lambda -
        \else\hbox{$\lambda^-_\tau$}\fi}